\documentclass[conference]{IEEEtran}

\usepackage{algorithm}
\makeatletter
\newcommand\fs@betterruled{%
  \def\@fs@cfont{\bfseries}\let\@fs@capt\floatc@ruled
  \def\@fs@pre{\vspace*{5pt}\hrule height.8pt depth0pt \kern2pt}%
  \def\@fs@post{\kern2pt\hrule\relax}%
  \def\@fs@mid{\kern2pt\hrule\kern2pt}%
  \let\@fs@iftopcapt\iftrue}
\floatstyle{betterruled}
\restylefloat{algorithm}
\makeatother
\usepackage{amsmath,graphicx}
\usepackage{amssymb}
\usepackage{algpseudocode}
\usepackage{cite}
\usepackage{amsfonts}
\usepackage{graphicx,subfigure}
\usepackage{fancyhdr}  %
\usepackage{cases}
\usepackage{extarrows}
\usepackage{multirow,tabularx}
\usepackage[english]{babel}
\newcommand{\squeezeup}{\vspace{-1mm}}

\newcommand{\squeezeupaaa}{\vspace{-0.5mm}}
\newcommand{\squeezedown}{\vspace{2mm}}

\newcommand{\n}{\mathbf{n}}

\newcommand{\s}{\mathbf{s}}
\renewcommand{\t}{\mathbf{t}}
\renewcommand{\u}{\mathbf{u}}
\renewcommand{\v}{\mathbf{v}}

\newcommand{\y}{\mathbf{y}}

\newcommand{\0}{\mathbf{0}}

\newcommand{\C}{\mathbf{C}}
\newcommand{\D}{\mathbf{D}}

\newcommand{\F}{\mathbf{F}}
\newcommand{\G}{\mathbf{G}}
\renewcommand{\H}{\mathbf{H}}

\renewcommand{\L}{\mathbf{L}}

\newcommand{\Q}{\mathbf{Q}}

\newcommand{\U}{\mathbf{U}}
\newcommand{\V}{\mathbf{V}}

\newcommand{\Real}{\mbox{$\mathbb{R}$}}
\newcommand{\Compl}{\mbox{$\mathbb{C}$}}

\newcommand{\tr}{\mathrm{tr}}

\IEEEoverridecommandlockouts   

\begin{document}

\title{Full Duplex Hybrid A/D Beamforming with Reduced Complexity Multi-Tap Analog Cancellation}
\author{
\IEEEauthorblockN{George~C.~Alexandropoulos$^1$, Md Atiqul Islam$^2$, and Besma Smida$^2$
 }
\IEEEauthorblockA{
$^1$Department of Informatics and Telecommunications, National and Kapodistrian University of Athens, Greece\\
$^2$Department of Electrical and Computer Engineering, University of Illinois at Chicago, USA\\
emails: alexandg@di.uoa.gr, \{mislam23, smida\}@uic.edu
}
\thanks{This work was partially funded by the National Science Foundation CAREER award \#1620902.}
}
\IEEEaftertitletext{\vspace{-4mm}}
\maketitle

\begin{abstract}
Although the hardware complexity of the analog self-interference canceller in full duplex Multiple Input Multiple Output (MIMO) designs does not necessarily scale with the number of transceiver antennas, exploiting the benefits of analog cancellation in massive MIMO systems with hundreds of antenna elements is still quite impractical. Hybrid Analog and Digital (A/D) beamforming architectures have been lately considered as a candidate technology for realizing massive MIMO transceivers with very large number of antenna elements, but with much fewer numbers of Radio Frequency (RF) chains. In this paper, we present a novel architecture for full duplex hybrid A/D beamforming transceivers including multi-tap analog cancellation with reduced number of taps and simple multiplexers for efficient signal routing among the transceiver RF chains. Capitalizing on the proposed transceiver architecture, we present a joint design of analog cancellation and A/D beamforming with the objective to maximize the achievable full duplex rate performance. Representative millimeter wave simulation results demonstrate the effectiveness of the proposed architecture and algorithmic framework for enabling simultaneous uplink and downlink communications with reduced complexity analog self-interference cancellation.
\end{abstract}


\section{Introduction}
In band full duplex, also known shortly as Full Duplex (FD), is a candidate technology for the Release $17$ of the fifth Generation (5G) New Radio (NR) standard enabling simultaneous UpLink (UL) and DownLink (DL) communication within the entire frequency band \cite{Sab14}. An FD radio can transmit and receive at the same time and frequency resource units, consequently, it can double the spectral efficiency achieved by a Half Duplex (HD) radio. Current wireless systems exploit Multiple Input Multiple Output (MIMO) communication, where increasing the number of Transmitter (TX) and Receiver (RX) antennas can increase the spatial Degrees of Freedom (DoF), hence boosting rate performance. Combining FD with MIMO operation can provide further spectral efficiency gains \cite{Rii11, Nguyen2013, Bha14, SofNull_2016, Tam2016, GA2016,GA2019}.

FD radios suffer from Self Interference (SI), a term referring to the signal transmitted by the FD radio TX that leaks to the FD radio RX. At the RX of the FD radio, the SI power can be many times stronger than the power of the received signal of interest. Consequently, SI can severely degrade the reception of the signal of interest, and thus SI mitigation is required in order to maximize the spectral efficiency gain of the FD operation. As the number of antennas increases, mitigating SI becomes more challenging, since more antennas naturally result in more SI components. Conventional SI suppression techniques in Single-Input Single-Output (SISO) systems include propagation domain isolation, analog domain suppression, and digital cancellation\cite{Bha14,korpi2014widely}. 
Although analog SI cancellation in FD MIMO systems can be implemented through SISO replication, its hardware requirements scale with the number of TX/RX antennas. The authors in \cite{Rii11,SofNull_2016} presented spatial suppression techniques that alleviate the need for analog SI cancellation, which was replaced solely by digital TX/RX beamforming. In \cite{alexandropoulos2017joint}, a joint design of multi-tap analog cancellation and TX/RX beamforming, where the number of taps does not scale with the product of TX and RX antenna elements, was proposed.
The FD technology has been lately theoretically combined with Hybrid analog and digital BeamForming (HBF) \cite{Molisch_HBF_2017} to enable simultaneous UL and DL communications in massive MIMO systems operating in the millimeter wave frequency bands \cite{xiao2017full,satyanarayana2018hybrid,cai2019robust,Vishwanath_2019}. These works mainly assigned the role of SI mitigation to the hybrid beamformers and/or deployed analog SI cancellation that scales with the number of TX/RX antennas. 

In this paper, we present a novel hardware architecture for FD HBF systems enabling the joint design of A/D TX/RX beamformers with reduced complexity tap-based analog cancellation. The proposed analog canceller interconnects a subset of the outputs of the TX Radio Frequency (RF) chains to a subset of the inputs to the RX RF chains in order to ensure that the residual SI signal after A/D TX precoding and analog RX combining remains below the noise floor. Our indicative simulation results with the proposed architecture and an example FD HBF algorithmic framework showcase a $1.7$ times rate improvement over HD HBF communication.      
\begin{figure*}
	\begin{center}
	\includegraphics[width=7in]{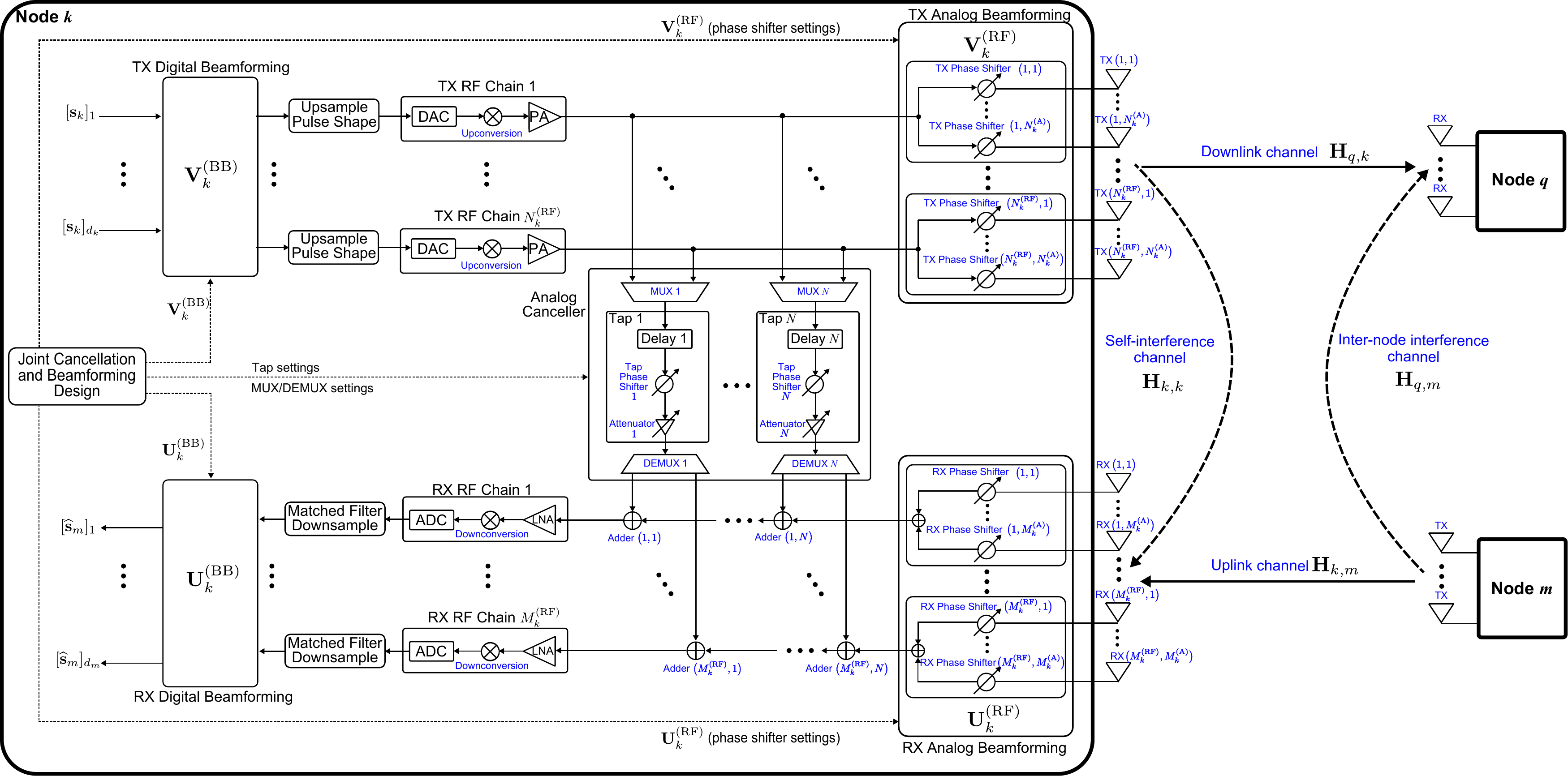}
	\caption{The considered bidirectional communication system with the proposed FD HBF architecture in the MIMO node $k$ including $N$-tap analog cancellation and A/D TX/RX beamforming. The HD multi-antenna nodes $q$ and $m$ communicate with node $k$ in the DL and UL directions, respectively. Each TX RF chain consists of a Digital to Analog Converter (DAC), a mixer upconverting the signal from BaseBand (BB) to RF, and a Power Amplifier (PA). A RX RF chain consists of a Low Noise Amplifier (LNA), a mixer downconverting the signal from RF to BB, and an Analog to Digital Converter (ADC). Upsampling and pulse shaping are used to prepare the BB signal for DAC and RF transmission at the TX side, whereas matched filtering and downsampling are used at the RX side before BB processing of the received RF signal.}
	\label{fig:FD_MIMO}
	\end{center}
	\squeezeup
	\squeezeup
	\squeezeup
\end{figure*}

\textit{Notation:} Vectors and matrices are denoted by boldface lowercase and boldface capital letters, respectively. The transpose and Hermitian transpose of $\mathbf{A}$ are denoted by $\mathbf{A}^{\rm T}$ and $\mathbf{A}^{\rm H}$, respectively, and $\det(\mathbf{A})$ is the determinant of $\mathbf{A}$, while $\mathbf{I}_{n}$ ($n\geq2$) is the $n\times n$ identity matrix and $\mathbf{0}_{n\times m}$ ($n,m\geq2$) is a $n\times m$ matrix with zeros. $\|\mathbf{A}\|_{\rm F}$ is the Frobenius norm of $\mathbf{A}$, $\|\mathbf{a}\|$ stands for $\mathbf{a}$'s Euclidean norm, and ${\rm diag}\{\mathbf{a}\}$ denotes a square diagonal matrix with $\mathbf{a}$'s elements in its main diagonal. $[\mathbf{A}]_{i,j}$ represents $\mathbf{A}$'s $(i,j)$-th element, while $[\mathbf{a}]_{i}$ denotes the $i$-th element of $\mathbf{a}$. $\Real$ and $\Compl$ represent the real and complex number sets, respectively, and $|\cdot|$ denotes the amplitude of a complex number.

\section{System Model and Proposed Architecture}  

\subsection{System Model}
We consider the $3$-user bidirectional communication system in Fig$.$~\ref{fig:FD_MIMO} comprising of a FD MIMO node $k$ equipped with $N_k$ TX and $M_k$ RX antenna elements, and two HD multi-antenna nodes $q$ and $m$ having $M_q$ and $N_m$ antennas, respectively. It is assumed that node $k$ communicates simultaneously (in the same time and frequency resources) with node $q$ in the DL and node $m$ in the UL. All nodes are considered capable of performing digital beamforming, which for simplicity we assume to be realized with linear filters. Node $k$ is also capable of analog TX/RX beamforming using the partially connected HBF architecture \cite{Molisch_HBF_2017}, as will be detailed in the sequel.  

It is assumed that node $m$ makes use of the digital precoding matrix $\V_m^{(\mathrm{BB})}\in\Compl^{N_m\times d_m}$ for processing in BaseBand (BB) its unit power symbol vector $\s_m\in\Compl^{d_m\times1}$ (chosen in practice from a discrete modulation set) before UL transmission. The dimension of $\s_m$ satisfies $d_m\leq\min\{M_{k}^{(\mathrm{RF})},N_m\}$ with $M_{k}^{(\mathrm{RF})}$ denoting the number of RX RF chains at node $k$. It holds $M_{k}^{(\mathrm{RF})}\leq M_{k}$, although in practical systems it can be $M_{k}^{(\mathrm{RF})}\ll M_{k}$. The constraint for $d_m$ certifies data decodability for the considered UL communication. It additionally holds that $\mathbb{E}\{\|\V_m^{(\mathrm{BB})}\s_m\|^2\}\leq {\rm P}_m$, where ${\rm P}_m$ is the total TX power of node $m$. On the DL, the reception node $q$ applies the digital combining matrix $\U_q^{(\mathrm{BB})}\in\Compl^{N_q\times d_k}$ in the BB received signal that includes the unit power symbol vector $\s_k\in\Compl^{d_k\times1}$ (again chosen from a discrete modulation set) transmitted from node $k$ such that $d_k\leq\min\{M_q,N_{k}^{(\mathrm{RF})}\}$ with $N_{k}^{(\mathrm{RF})}$ ($N_{k}^{(\mathrm{RF})}\leq N_{k}$, but practically it can be $N_{k}^{(\mathrm{RF})}\ll N_{k}$) denoting the number of TX RF chains at node $k$. Similarly, the latter constraint verifies the spatial DoF of the effective $M_q\times N_{k}^{(\mathrm{RF})}$ DL MIMO channel between the TX RF chains of node $k$ and the RX RF chains of node $q$. It is noted that each antenna at node $q$ is connected to a dedicated RF chain.
\squeezeup
\subsection{Proposed FD HBF Hardware Architecture} 
The proposed FD HBF hardware architecture, comprising of $N$-tap analog cancellation for the SI signal as well as A/D precoding and combining for the outgoing and incoming signals, is adopted for the MIMO node $k$, as depicted in the left part of Fig$.$~\ref{fig:FD_MIMO}. Differently from \cite{alexandropoulos2017joint}'s architecture that considered only fully digital TX/RX beamforming, node $k$ is capable of HBF through its partially connected beamforming architecture. As shown in the figure, the analog canceller interconnects the $N_{k}^{(\mathrm{RF})}$ inputs of the analog TX precoder to the $M_{k}^{(\mathrm{RF})}$ outputs of the analog RX combiner. The complexity of the analog canceller expressed in the number of taps $N$ is independent of the numbers $N_{k}$ and $M_{k}$ of the TX and RX antennas, respectively, and as it will be shown next, it scales with the product $\xi N_{k}^{(\mathrm{RF})}M_{k}^{(\mathrm{RF})}$ with $\xi<1$. This is in contrast to \cite{alexandropoulos2017joint} where the analog canceller interconnects the $N_{k}$ TX antenna inputs to the $M_{k}$ RX antenna outputs. 

\subsubsection{Partially Connected HBF} 
Each of the $N_{k}^{(\mathrm{RF})}$ TX RF chains of node $k$ is connected to a separate subset of the available TX antenna elements. As shown in Fig$.$~\ref{fig:FD_MIMO}, the $i$-th TX RF chain with $i=1,2,\ldots,N_{k}^{(\mathrm{RF})}$ is connected via phase shifters with $N_k^{({\rm A})}$ TX antenna elements, each denoted as ${\rm TX}(i,j)$ $\forall$$j=1,2,\ldots,N_k^{({\rm A})}$. Clearly, it holds $N_k=N_{k}^{(\mathrm{RF})}N_k^{({\rm A})}$ for the total number of TX antennas at node $k$. Stacking the values of the $N_k^{({\rm A})}$ phase shifters that connect each $i$-th TX RF chain with its antenna elements in a complex-valued $N_{k}^{({\rm A})}\times1$ vector $\v_i$, we can formulate the complex-valued $N_{k}\times N_{k}^{(\mathrm{RF})}$ analog TX precoder, as follows:
\begin{equation}\label{eq:TX_analog_precoder}
\mathbf{V}_k^{(\mathrm{RF})} = \left[ \begin{matrix}
\mathbf{v}_1  &  \mathbf{0}_{N_k^{({\rm A})}\times1}    &  \cdots    &  \mathbf{0}_{N_k^{({\rm A})}\times1}  \\
\mathbf{0}_{N_k^{({\rm A})}\times1}    &  \mathbf{v}_2  &  \cdots    &  \mathbf{0}_{N_k^{({\rm A})}\times1}  \\
\vdots   &  \vdots   &  \ddots    &	 \vdots  \\
\mathbf{0}_{N_k^{({\rm A})}\times1}    &  \mathbf{0}_{N_k^{({\rm A})}\times1}    &  \cdots    &  \mathbf{v}_{N_k^{({\rm RF})}}
\end{matrix}
\right].
\end{equation}
The elements of each $\v_i$ are assumed to have constant magnitude, i$.$e$.$, $|[\mathbf{v}_i]_{n}|^2=1/N_k^{({\rm A})}$ $\forall$$n=1,2,\ldots,N_k^{({\rm A})}$. We also assume that $\mathbf{v}_i\in\mathbb{F}_{\rm TX}$ $\forall$$i=1,2,\ldots,N_{k}^{(\mathrm{RF})}$, which means that all analog TX precoding vectors belong in a predefined beam codebook $\mathbb{F}_{\rm TX}$ including ${\rm card}(\mathbb{F}_{\rm TX})$ distinct vectors (or analog beams). Apart from applying $\mathbf{V}_k^{(\mathrm{RF})}$ in the analog domain to the signal before transmission, the symbol vector $\s_k$ is also processed in BB with the digital TX precoder $\V_k^{(\mathrm{BB})}\in\Compl^{N_{k}^{(\mathrm{RF})}\times d_k}$ (recall that $d_k\leq\min\{M_q,N_{k}^{(\mathrm{RF})}\}$) before entering into the $N_{k}^{(\mathrm{RF})}$ TX RF chains, as shown in Fig$.$~\ref{fig:FD_MIMO}. Similar to the UL communication from node $m$ to $k$, we assume that the DL transmission from node $k$ to $q$ is power limited according to $\mathbb{E}\{\|\V_k^{(\mathrm{RF})}\V_k^{(\mathrm{BB})}\s_k\|^2\}\leq {\rm P}_k$ with ${\rm P}_k$ being the total available TX power at node $k$. 

The RX of node $k$ is composed of an analog combiner connecting the RX antenna elements to the inputs of the RX RF chains, and a digital combiner that processes the outputs of the RX RF chains in BB before signal decoding. In particular, the $n$-th RX RF chain with $n=1,2,\ldots,M_{k}^{(\mathrm{RF})}$ is connected through phase shifters with $M_k^{({\rm A})}$ distinct RX antennas; these phase shifters are denoted as ${\rm RX}(n,\ell)$ $\forall$$\ell=1,2,\ldots,M_k^{({\rm A})}$. It should hold that $M_k=M_{k}^{(\mathrm{RF})}M_k^{({\rm A})}$ for the total number of RX antennas at node $k$. We define the complex-valued $M_{k}\times M_{k}^{(\mathrm{RF})}$ analog RX combiner $\mathbf{U}_k^{(\mathrm{RF})}$ having a similar block diagonal structure to \eqref{eq:TX_analog_precoder}. In particular, $\mathbf{U}_k^{(\mathrm{RF})}$ contains $\u_n$'s with $n=1,2,\ldots,M_{k}^{(\mathrm{RF})}$ in the diagonal,       
where each $\u_n$ contains the constant magnitude values of the $M_k^{({\rm A})}$ phase shifters (i$.$e$.$, $|[\mathbf{u}_n]_{j}|^2=1/M_k^{({\rm A})}$ $\forall$$j=1,2,\ldots,M_k^{({\rm A})}$) connecting each $n$-th RX RF chain with its antenna elements. We also assume that $\mathbf{u}_n\in\mathbb{F}_{\rm RX}$ $\forall$$n$, i.e., all analog RX combiners belong in a predefined beam codebook $\mathbb{F}_{\rm RX}$ having ${\rm card}(\mathbb{F}_{\rm RX})$ vectors. Finally, $\U_k^{(\mathrm{BB})}\in\Compl^{M_{k}^{(\mathrm{RF})}\times d_m}$ with $d_m\leq\min\{M_{k}^{(\mathrm{RF})},N_m\}$ represents the digital RX combiner at node $k$. 
 
\subsubsection{Multi-Tap Analog Cancellation} 
The analog canceller at node $k$ consists of $N$ taps with each tap connected via a $N_k^{\mathrm{(RF)}}$-to-$1$ MUltipleXer (MUX) to all $N_k^{\mathrm{(RF)}}$ outputs of the respective TX RF chains. A tap includes a fixed delay, a variable phase shifter, and a variable attenuator \cite{Kol16,alexandropoulos2017joint}. To route the cancellation signal to one of the adders located just before the RX RF chains, the output of each tap is connected to a $1$-to-$M_k^{\mathrm{(RF)}}$ DEMUltipleXer (DEMUX). There is a total of $NM_k^{\mathrm{(RF)}}$ such adders and we use the notation ``Adder ($i,j$)'' to label the adder that connects DEMUX $j$ to RX RF chain $i$, where $i=1,2,\ldots,M_k^{\mathrm{(RF)}}$ and $j=1,2,\ldots,N$. The adders before the RX RF chains can be implemented via power combiners or directional couplers, while the analog RF MUXs/DEMUXs can be implemented with RF switches. Clearly, the proposed analog canceller interconnects the outputs of some of the available TX RF chains to the inputs of some of the RX RF chains, and in contrast to \cite{alexandropoulos2017joint}, the size of each MUX/DEMUX depends on the number of TX/RF RF chains and not on the number of TX/RX antennas. Similar to \cite{alexandropoulos2017joint}, we model in BB the analog processing realized by the analog canceller as $\C_k\triangleq \L_3\L_2\L_1\in\Compl^{M_k^{\mathrm{(RF)}}\times N_k^{\mathrm{(RF)}}}$, where $\L_1\in\Real^{N \times N_k^{\mathrm{(RF)}}}$, $\L_2\in \Compl^{N\times N}$, and $\L_3\in\Real^{M_k^{\mathrm{(RF)}}\times N}$. The elements $[\L_1]_{j,\ell}$ and $[\L_3]_{i,j}$ with $j=1,2,\ldots,N$, $\ell=1,2,\ldots,N_k^{\mathrm{(RF)}}$, and $i=1,2,\ldots,M_k^{\mathrm{(RF)}}$ take the binary values $0$ or $1$, and it must hold that 
\begin{equation}\label{Eq:L_1_L_3}
\sum_{\ell=1}^{N_k^{\mathrm{(RF)}}}[\L_1]_{j,\ell} = \sum_{i=1}^{M_k^{\mathrm{(RF)}}}[\L_3]_{i,j} = 1 \,\,\,\forall j=1,2,\ldots,N.
\end{equation}
The $\L_2$ in $\C_{k}$ is a diagonal matrix whose complex entries represent the attenuation and phase shift of the canceller taps; the magnitude and phase of the element $[\L_2]_{i,i}$ with $i=1,2,\ldots,N$ specify the attenuation and phase of the $i$-th tap. Recall that the tap delays in each canceller tap are fixed, hence, we model the effects of the $i$-th tap delay as a phase shift that is incorporated to the phase of $[\L_2]_{i,i}$. 

\squeezeup
\subsection{Received Signal Models} 
Using the previously described system configuration, the BB received signal $\y_{q}\in\Compl^{M_q\times 1}$ at node $q$ in the DL communication can be mathematically expressed as
\begin{equation}\label{Eq:Received_q}
\y_{q} \triangleq \H_{q,k}\V_k^{(\mathrm{RF})}\V_k^{(\mathrm{BB})}\s_k + \H_{q,m}\V_{m}\s_{m} + \n_{q}, 
\end{equation}
where $\H_{q,k}\in\Compl^{M_q\times N_k}$ is the DL channel gain matrix (i$.$e$.$, between nodes $q$ and $k$), $\H_{q,m}\in\Compl^{M_q\times N_m}$ denotes the channel gain matrix for inter-node interference (i$.$e$.$, between nodes $q$ and $m$), and $\n_{q}\in\Compl^{M_q\times 1}$ represents the Additive White Gaussian Noise (AWGN) at node $q$ with variance $\sigma_{q}^{2}$. In the UL communication, the symbol vector $\hat{\s}_m\in\Compl^{d_m\times 1}$ used for the estimation of $\s_m$ at the FD HBF node $k$ is derived as
\squeezedown
\begin{align}\label{Eq:Estimated_m} 
\nonumber\hat{\s}_m \triangleq&  \left(\U_k^{(\mathrm{BB})}\right)^{\rm H}\left(\left(\U_k^{(\mathrm{RF})}\right)^{\rm H}\H_{k,k}\V_k^{(\mathrm{RF})}+\C_k\right)\V_k^{(\mathrm{BB})}\s_k\\
&+ \left(\U_k^{(\mathrm{BB})}\right)^{\rm H}\left(\U_k^{(\mathrm{RF})}\right)^{\rm H}\left(\H_{k,m}\V_{m}\s_{m} + \n_k\right),
\end{align}
\squeezedown
where $\H_{k,k}\in\Compl^{M_k\times N_k}$ denotes the SI channel seen at the RX antennas of node $k$ due to its own DL transmission, $\H_{k,m}\in\Compl^{M_k\times N_m}$ is the UL channel gain matrix (i$.$e$.$, between nodes $k$ and $m$), and $\n_{k}\in\Compl^{M_k\times 1}$ denotes the received AWGN at node $k$ with variance $\sigma_{k}^{2}$. The first term in \eqref{Eq:Estimated_m} describes the residual SI signal after analog cancellation and A/D TX/RX beamforming, while its second term contains the A/D RX combined signal transmitted from node $m$ plus AWGN. In contrast to \cite{alexandropoulos2017joint}, $\C_k$ needs to cancel the SI channel $(\U_k^{(\mathrm{RF})})^{\rm H}\H_{k,k}\V_k^{(\mathrm{RF})}$, which is a matrix of dimension $M_k^{\mathrm{(RF)}}\times N_k^{\mathrm{(RF)}}$ and not the actual $M_k\times N_k$ SI channel $\H_{k,k}$. 

\section{Joint Design Problem Formulation}\label{sec:Design}
We focus on the FD HBF node $k$ in Fig$.$~\ref{fig:FD_MIMO} and present a sum-rate optimization framework for the joint design of $\C_k$, $\V_k^{(\mathrm{RF})}$, $\V_k^{(\mathrm{BB})}$, $\U_k^{(\mathrm{RF})}$, and $\U_k^{(\mathrm{BB})}$. Using the notation $\V_k\triangleq\V_k^{(\mathrm{RF})}\V_k^{(\mathrm{BB})}$ and assuming Gaussian signaling and capacity-achieving combining at node $q$, the achievable DL rate that is a function of the A/D TX precoding matrices $\V_k^{(\mathrm{RF})}$ and $\V_k^{(\mathrm{BB})}$ of node $k$ as well as the digital TX precoder $\V_m$ of node $m$, is given by
\begin{equation}\label{eq:DL_Rate}
\mathcal{R}_{\rm DL} = \log_2\left({\rm det}\left(\mathbf{I}_{M_q}+\H_{q,k}\V_k\V_k^{\rm H}\H_{q,k}^{\rm H}\Q_{q}^{-1}\right)\right),
\end{equation}
\squeezeup
where $\Q_{q}\in\mathbb{C}^{M_q\times M_q}$ denotes the covariance matrix of the Interference-plus-Noise (IpN) at node $q$ that is obtained as
\begin{equation}\label{eq:Interference_q}
\Q_{q} \triangleq \H_{q,m}\V_{m}\V_{m}^{\rm H}\H_{q,m}^{\rm H} + \sigma_q^2\mathbf{I}_{M_q}.
\end{equation}
We hereinafter assume that there is no inter-node interference between the HD multi-antenna nodes $q$ and $m$ due to, for example, appropriate node scheduling \cite{GA2016} for the FD operation at node $k$. The latter assumption translates to setting the channel matrix between those involved nodes in \eqref{Eq:Received_q} as $\H_{q,m}=\0_{M_q\times N_k}$, which means that \eqref{eq:Interference_q} simplifies to $\Q_{q}=\sigma_q^2\mathbf{I}_{M_q}$.

For the computation of the achievable UL rate, we use the notation $\U_k\triangleq\U_k^{(\mathrm{RF})}\U_k^{(\mathrm{BB})}$ to express this rate as a function of the A/D RX combiners $\U_k^{(\mathrm{RF})}$ and $\U_k^{(\mathrm{BB})}$, the A/D TX precoders $\V_k^{(\mathrm{RF})}$ and $\V_k^{(\mathrm{BB})}$, and the analog cancellation matrix $\C_k$ of node $k$ as well as of the digital TX precoder $\V_m$ of node $m$. Using \eqref{Eq:Estimated_m}, the UL rate is given by
\begin{equation}\label{eq:UL_Rate}
\mathcal{R}_{\rm UL} = \log_2\left({\rm det}\left(\mathbf{I}_{d_m}+\U_k^{\rm H}\H_{k,m}\V_m\V_m^{\rm H}\H_{k,m}^{\rm H}\U_k\Q_{k}^{-1}\right)\right).
\end{equation}
where $\Q_{k}\in\mathbb{C}^{d_m\times d_m}$ denotes the IpN covariance matrix after A/D RX combining at node $q$, which can be expressed as
\begin{equation}\label{eq:Interference_k}
\begin{split}
\Q_{k} \triangleq& \left(\U_k^{(\mathrm{BB})}\right)^{\rm H}\tilde{\mathbf{H}}_{k,k}\V_k^{(\mathrm{BB})}\left(\V_k^{(\mathrm{BB})}\right)^{\rm H}\tilde{\mathbf{H}}_{k,k}^{\rm H}\U_k^{(\mathrm{BB})} \\&
+ \sigma_k^2\left(\U_k^{(\mathrm{BB})}\right)^{\rm H}\left(\U_k^{(\mathrm{RF})}\right)^{\rm H}\U_k^{(\mathrm{RF})}\U_k^{(\mathrm{BB})}.
\end{split}
\end{equation}
In the latter expression, $\tilde{\mathbf{H}}_{k,k}\in\Compl^{M_k^{\mathrm{(RF)}}\times N_k^{\mathrm{(RF)}}}$ denotes the effective SI channel after performing analog TX/RX beamforming and analog cancellation, which is defined as
\begin{equation}\label{eq:New_Matrix}
\tilde{\mathbf{H}}_{k,k} \triangleq \left(\U_k^{(\mathrm{RF})}\right)^{\rm H}\H_{k,k}\V_k^{(\mathrm{RF})}+\C_k.
\end{equation}
Using the expressions \eqref{eq:DL_Rate} and \eqref{eq:UL_Rate} for the achievable DL and UL rates, respectively, the sum-rate optimization problem for the joint design of the analog canceller and the A/D TX/RX beamformers is mathematically expressed as
\begin{equation*}\label{eq:optim}
\begin{split}
  &\mathcal{OP}: \max_{\C_k,\V_k^{(\mathrm{RF})},\V_k^{(\mathrm{BB})},\U_k^{(\mathrm{RF})},\U_k^{(\mathrm{BB})}} \mathcal{R}_{\rm DL} + \mathcal{R}_{\rm UL}
	\\& \textrm{s.t.}~~\tr\{\V_k^{(\mathrm{RF})}\V_k^{(\mathrm{BB})}\left(\V_k^{(\mathrm{BB})}\right)^{\rm H}\left(\V_k^{(\mathrm{RF})}\right)^{\rm H}\}\leq{\rm P}_k,\,\,({\rm C1})
	\\& \C_k = \L_3\L_2\L_1 \,\,{\rm with}\,\,\eqref{Eq:L_1_L_3}\,\,{\rm and}\,\,[\L_2]_{i,j}=0\,\,{\rm for}\,\,i\neq j,\hspace{0.21cm}({\rm C2}) 
	\\& \left\|\left[\tilde{\mathbf{H}}_{k,k}\V_k^{(\mathrm{BB})}\right]_{(j,:)}\right\|^2 \leq \rho_{\rm A}\,\,\forall j=1,2,\ldots,M_k^{\mathrm{(RF)}},\hspace{0.175cm}({\rm C3})
	\\& \mathbf{u}_j\in\mathbb{F}_{\rm RX}\,\,\forall j\,\,{\rm and}\,\, \mathbf{v}_n\in\mathbb{F}_{\rm TX}\,\,\forall n=1,2,\ldots,N_k^{({\rm RF})},({\rm C4})
\end{split}
\end{equation*}
where constraint $({\rm C1})$ relates to the average TX power at node $k$ and constraint $({\rm C2})$ refers to the hardware capabilities of the analog canceller. Constraint $({\rm C3})$ imposes the threshold $\rho_{\rm A}\in\Real$ on the average power of the residual SI signal after analog cancellation and analog TX/RX beamforming. 
Finally, constraint $({\rm C4})$ refers to the predefined TX and RX beam codebooks. To tackle $\mathcal{OP}$, which is a nonconvex problem with nonconvex constraints, we adopt a similar to \cite{alexandropoulos2017joint} decoupled way that in this case requires at most $\alpha_{\max}\triangleq N_{k}^{(\mathrm{RF})}-1$ iterations including closed form expressions for the design parameters. We first solve for $\C_k$, $\V_k^{(\mathrm{RF})}$, $\V_k^{(\mathrm{BB})}$, and $\U_k^{(\mathrm{RF})}$ maximizing the DL rate, and then find $\U_k^{(\mathrm{BB})}$ maximizing the UL rate. Specifically, we formulate the following optimization subproblem for the design of $\C_k$, $\V_k^{(\mathrm{RF})}$, $\V_k^{(\mathrm{BB})}$, and $\U_k^{(\mathrm{RF})}$: 
\begin{equation*}\label{eq:optim1}
\begin{split}
  \mathcal{OP}1: &\max_{\C_k,\V_k^{(\mathrm{RF})},\V_k^{(\mathrm{BB})},\U_k^{(\mathrm{RF})}} \mathcal{R}_{\rm DL}~~\textrm{s.t.}~~({\rm C1}),\,({\rm C2}),\,\text{and}\,({\rm C3}).
\end{split}
\end{equation*}
\begin{algorithm}[t!]\caption{Digital TX Precoder Design}\label{DL_Precoding}
\begin{algorithmic}[1]
\Statex \textbf{Input:} ${\rm P}_k$, $\V_k^{(\mathrm{RF})}$ and $\U_k^{(\mathrm{RF})}$ solving $\mathcal{OP}2$, $\H_{k,k}$, and $\H_{q,k}$ as well as a realization of $\C_k$ for a given $N$ satisfying constraint $({\rm C2})$.
\State Set $\tilde{\mathbf{H}}_{k,k} = \left(\U_k^{(\mathrm{RF})}\right)^{\rm H}\H_{k,k}\V_k^{(\mathrm{RF})}+\C_k$. 
\State Obtain $\D_k$ with the $N_k^{\mathrm{(RF)}}$ right-singular vectors of $\widetilde{\H}_{k,k}$ corresponding to the singular values in descending order.
\For{$\alpha=\alpha_{\max},\alpha_{\max}-1,\ldots,2$}
		\State Set $\F_k=[\D_k]_{(:,N_k^{\mathrm{(RF)}}-\alpha+1:N_k^{\mathrm{(RF)}})}$.
		\State Set $\G_k$ as the optimum precoding for the effective
		       \Statex \hspace{0.55cm}DL MIMO channel $\H_{q,k}\V_k^{(\mathrm{RF})}\F_k$ given ${\rm P}_k$.
		\If{$\|[\widetilde{\H}_{k,k}\F_k\G_k]_{(i,:)}\|^2\leq\rho_{\rm A}$ $\forall i=1,\ldots,M_k^{\mathrm{(RF)}}$,}
			 \State Output $\V_k^{(\mathrm{BB})}=\F_k\G_k$ and stop the algorithm.
		\EndIf
\EndFor
\State Set $\F_k=[\D_k]_{(:,N_k^{\mathrm{(RF)}})}$ and $\G_k={\rm P}_k^{1/2}$.
\If{$|[\widetilde{\H}_{k,k}\F_k\G_k]_{i}|^2\leq\rho_{\rm A}$ $\forall i=1,\ldots,M_k^{\mathrm{(RF)}}$,}
			 \State Output $\V_k^{(\mathrm{BB})}=\F_k\G_k$ and stop the algorithm.
\Else
			 \State Output that the $\C_k$ realization does not meet 
			 \Statex \hspace{0.55cm}the residual SI constraint.
\EndIf
\end{algorithmic}
\end{algorithm}
For the solution of the latter problem we use an alternating optimization approach. First, we find $\V_k^{(\mathrm{RF})}$ and $\U_k^{(\mathrm{RF})}$ constrained as $({\rm C4})$ that boost the DL rate, while minimizing the SI signal before any other form of cancellation. Particularly, we perform the following exhaustive search:
\setlength{\abovedisplayskip}{3.5pt}
\setlength{\belowdisplayskip}{3.5pt}
\begin{equation*}\label{eq:optim2}
\begin{split}
  \mathcal{OP}2: &\max_{\V_k^{(\mathrm{RF})},\U_k^{(\mathrm{RF})}} \frac{\left\|\H_{q,k}\V_k^{(\mathrm{RF})}\right\|_{\rm F}}{\left\|\left(\U_k^{(\mathrm{RF})}\right)^{\rm H}\H_{k,k}\V_k^{(\mathrm{RF})}\right\|_{\rm F}}~~\textrm{s.t.}~~({\rm C4}).
\end{split}
\end{equation*}
In the sequel, given the solution for $\mathcal{OP}2$ and supposing that the available number of analog canceller taps $N$ and a realization of $\C_k$ satisfying $({\rm C2})$ are given, we seek for $\V_k^{(\mathrm{BB})}$ maximizing the DL rate while meeting $({\rm C1})$ and $({\rm C3})$. The latter procedure is repeated for all allowable realizations of $\C_k$ for the given $N$ in order to find the best $\V_k^{(\mathrm{BB})}$ solving $\mathcal{OP}1$; this procedure is summarized in Algorithm~\ref{DL_Precoding}. 
The values for $\C_k$, $\V_k^{(\mathrm{RF})}$, $\V_k^{(\mathrm{BB})}$, and $\U_k^{(\mathrm{RF})}$ solving $\mathcal{OP}1$ are finally substituted into the achievable UL rate expression  $\mathcal{R}_{\rm UL}$ in \eqref{eq:UL_Rate}. The $\U_k^{(\mathrm{BB})}$ maximizing this point-to-point MIMO rate is obtained in closed form using \cite[Sec. 4.2]{J:George_Elsevier_13}.

\squeezeupaaa
\section{Simulation Results and Discussion}
In this section, we investigate the performance of the considered $3$-user bidirectional communication system for the case where $N_{k}^{(\mathrm{RF})} = M_{q} = 4$, $M_{k}^{(\mathrm{RF})}=2$, $N_{m}=1$, and $N_{k}^{(\mathrm{A})} = M_{k}^{(\mathrm{A})} = \{16, 128\}$. We have assumed that the antennas at the FD HBF node $k$ are arranged in Uniform Linear Arrays (ULAs) with $\lambda/2$ space separation between adjacent elements, where $\lambda$ is the wavelength. The distance and angle between the TX and RX ULAs at node $k$ were set respectively to $d = 2\lambda$ and $\omega = \pi/6$ \cite{xiao2017full}. The DL and UL channels were simulated as millimeter wave clustered channels, as described in \cite[eq$.$ (2)]{satyanarayana2018hybrid} with pathloss $110$dB. In addition, the SI channel was modeled as Rician \cite[eq. (7)]{satyanarayana2018hybrid} with $K$-factor $35$dB and pathloss $40$dB. The RX noise floors at both nodes $k$ and $q$ were set to $-110$dBm, resulting in an effective dynamic range of $62$dB for $14$-bit ADCs with a $10$dB peak-to-average power ratio. Hence, to avoid saturation, the residual SI power after analog cancellation at the input of each RX RF chain needs to be below $-47$dBm. Non-ideal $N$-tap analog cancellation has been considered as in \cite{alexandropoulos2017joint}, and for the analog TX/RX beamformers, we have used beam codebooks based on the discrete Fourier transform  matrix. 

\begin{figure}[!tpb]
	\begin{center}
	\includegraphics[width=0.48\textwidth]{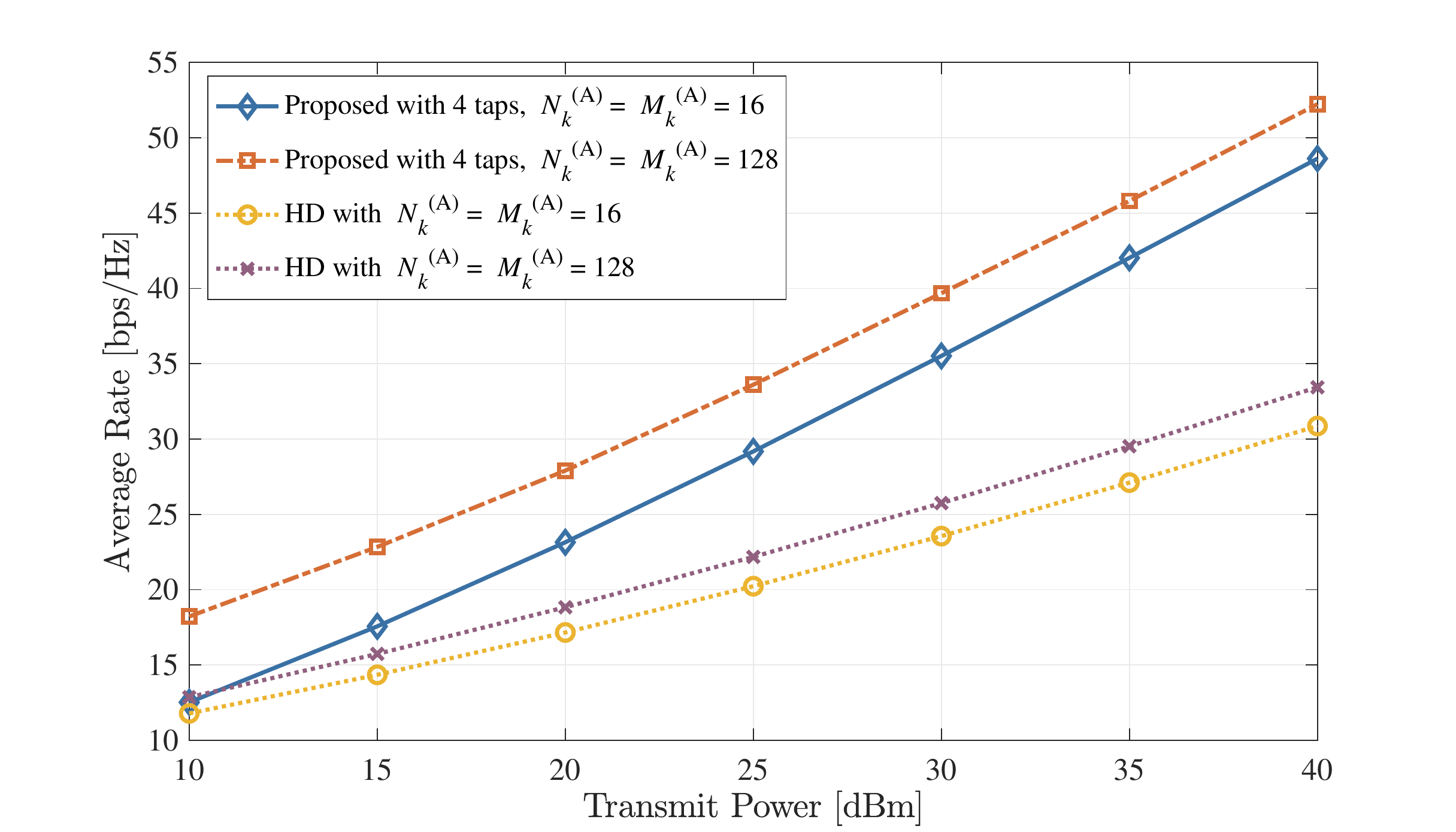}
	\caption{Average FD and HD rates vs the transmit powers in dBm for $N_{k}^{(\mathrm{RF})} = M_{q} = N = 4$, $M_{k}^{(\mathrm{RF})}=2$, $N_{m}=1$, and different values for $N_{k}^{(\mathrm{A})}$ and $M_{k}^{(\mathrm{A})}$ at node $k$.}
	\label{fig:FD_rate}
	\end{center}
\end{figure}

The achievable FD rate as a function of the transmit powers of nodes $k$ and $m$ is illustrated in Fig$.$~\ref{fig:FD_rate} for $N=4$ taps for the proposed analog canceller, which translates to $50\%$ reduction in the number of taps compared to a case that connects all outputs of the TX RF chains to every input to the RX RF chains. For the rate results, we averaged over $1000$ independent channel realizations and calculated the FD rate with the proposed Algorithm~\ref{DL_Precoding}, as well as the achievable HD rate. It is shown in the figure that all rates increase with increasing transmit power, and no rate saturation is exhibited for the proposed FD HBF technique. The latter trend witnesses the effective adaptability of Algorithm~\ref{DL_Precoding} to the SI conditions for both considered pairs of $N_{k}^{(\mathrm{A})}$ and $M_{k}^{(\mathrm{A})}$. As an indicative example, it is shown that for $40$dBm transmit powers, the proposed approach results in a $52$bps/Hz achievable rate, which is around $1.7$ times more than the achievable HD rate.



\bibliographystyle{IEEEtran}
\bibliography{refs}
\end{document}